\documentclass[aps,pra,twocolumn,amsfonts,amssymb,amsmath,showpacs,
floatfix,nofootinbib,groupedaddress,citesort]{revtex4-1}
\usepackage{mathrsfs}
\usepackage{amsfonts}
\usepackage{amstext}
\usepackage{amsmath}
\usepackage{amssymb}
\usepackage{bm}
\usepackage{bbm}
\usepackage[dvips]{graphicx}
\def\qed{\leavevmode\unskip\penalty9999 \hbox{}\nobreak\hfill
     \quad\hbox{\leavevmode  \hbox to.77778em{%
              \hfil\vrule   \vbox to.675em%
               {\hrule width.6em\vfil\hrule}\vrule\hfil}}
     \par\vskip3pt}

\begin{document}

\title{Constructing UMEB from maximally entangled basis  
\\}

\author{Yu Guo}
\affiliation{School of Mathematics and Computer Science, Shanxi Datong University, Datong, Shanxi 037009, China}%




\begin{abstract}

A new way of constructing unextendible maximally entangled basis (UMEB)
from maximally entangled basis (MEB) is proposed.
Consequently, it is shown that if there is an $N$-member UMEB in $\mathbb{C}^d\otimes \mathbb{C}^d$,
then there exists a $(qd)^2-q(d^2-N)$-member UMEB in $\mathbb{C}^{qd}\otimes \mathbb{C}^{qd}$ for any $q\in\mathbb{N}$.
This improves the results in [Phys. Rev. A 90, 034301(2014)], which shows that
there exists a $(qd)^2-(d^2-N)$-member UMEB in $\mathbb{C}^{qd}\otimes \mathbb{C}^{qd}$ provided that
an $N$-member UMEB exsits in $\mathbb{C}^d\otimes \mathbb{C}^d$.
In addition, a very easy way of constructing UMEB in $\mathbb{C}^d\otimes \mathbb{C}^{d'}$ with $d<d'$ is presented,
which promotes and covers all the previous related work.

\end{abstract}

\maketitle

\section{Introduction}

Entanglement as a fundamental feature
of quantum physics has been proved to be a central resource in quantum information
and quantum computation \cite{Horodecki2009,Guhne}.
One of the important problems in this field is
to characterize entanglement not only physically but also mathematically.
Consequently, the unextendible bases have attracted much attention in recent years
\cite{Bennett1999,Bravyi,Guowu2014,Pittenger,Fei2015,DiVincenzo,Bennett1999pra,
Leinaas2011pra,Skowronek2011,
Bravyi2004,Terhal2001,Terhal2001,Augusiak,Duan2010pra,
Horodecki2003prl,Alon,Feng2006,Chen2014,Johnston2014,Chen2014jmp,
Chakrabarty,Nan2015,Zhang2016ijtp,Nizamidin2015,Chen,
Limaosheng,Wangyanling,Guoyu2015jpa,Guoyu2015qip,Wangyanling2016}.
The first unextendible basis is
the unextendible product basis (UPB) \cite{Bennett1999},
which indicates nonlocality without
entanglement since it can not be distinguished by local
measurements and classical communication \cite{Bennett1999}.
Moreover, it can be used for
constructing bound entangled states \cite{Pittenger,DiVincenzo,Leinaas2011pra,Skowronek2011,Bravyi2004}.
Later,
unextendible maximally entangled basis (UMEB)~\cite{Bravyi}
and the unextendible entangled basis with Schmidt
number $k$ (UEBk)~\cite{Guowu2014} were investigated extensively.

The highest priority problem in studying the unextendible basis is to check
whether they exist in the given bipartite state space.
The existence problem of UPB and UEBk have been resolved completely.
The UPB exists in any $\mathbb{C}^d\otimes \mathbb{C}^{d'}$ with
$d\neq2$ and $d'\neq2$ \cite{Bennett1999,Chen2014} and
there exists UEBk in $\mathbb{C}^d\otimes \mathbb{C}^{d'}$
for any $d$ and $d'$ \cite{Guowu2014,Guoyu2015qip}.
Although considerable progress has been made,
the existence of UMEB still remains open.
It has been shown that there is no UMEB in the two-qubit system, a
6-member UMEB exists in $\mathbb{C}^3\otimes \mathbb{C}^3$ and
a 12-member UMEB exists in $\mathbb{C}^4\otimes \mathbb{C}^4$~\cite{Bravyi}.
Later, B. Chen and S.-M. Fei proved in Ref. \cite{Chen} that
there exists a set of $d^2$-member UMEB in
$\mathbb{C}^d\otimes \mathbb{C}^{d'}$ ($\frac{d'}{2}<d<d'$) and
questioned the existence of UMEBs in the case of $d'\geq 2d$.
Ref. \cite{Limaosheng} proved that there might
be two sets of UMEBs in $\mathbb{C}^d\otimes \mathbb{C}^{d'}$ whenever $d\neq d'$, and
an explicit construction of UMEBs is put forward.
A 30-member UMEB in $\mathbb{C}^6\otimes \mathbb{C}^6$
was given in Ref.~\cite{Wangyanling} and they
give a method of constructing UMEB in $\mathbb{C}^{qd}\otimes \mathbb{C}^{qd}$ from that of
$\mathbb{C}^d\otimes \mathbb{C}^d$.
Recently, Wang et al. proved that for any $d$ there exists a
UMEB except $d=p$ or $2p$, where $p\equiv3$ mod 4
and $p$ is a prime \cite{Wangyanling2016}.
They also presented a 23-member UMEB in $\mathbb{C}^5\otimes \mathbb{C}^5$
and a $45$-member UMEB in $\mathbb{C}^7\otimes \mathbb{C}^7$.

Here, we propose a new scenario of constructing
UMEB via the space decomposition.
Our method improves all the previous work
about UMEB.
The rest of this paper is constructed as follows. In
Sec. II, we introduce some related notations and terminologies.
Sec. III deals with the case of $d=d'$ and
Sec. IV discusses the case of $d<d'$.
We give an easy method to get a $(qd)^2-q(d^2-N)$-member UMEB
in $\mathbb{C}^{qd}\otimes \mathbb{C}^{qd}$ from the MEB and
UMEB in $\mathbb{C}^d\otimes \mathbb{C}^d$.
It has much less members comparing with the one in \cite{Wangyanling} and
the subspace structure of the associated Hilbert-Schmidt space depicts
the formation of UMEB in a concise way.
For the case of $d\neq d'$, we settled down the constructing of UMEB
based on a more completely rational scenario.
We conclude in Sec. V at last.

\section{Definition and preliminary}

Throughout this paper, we always assume that $d\leq d'$.
A state $|\psi\rangle\in\mathbb{C}^d\otimes\mathbb{C}^{d'}$ is
called a maximally entangled state if
it can be
written as $|\psi\rangle=\frac{1}{\sqrt{d}}\sum_{i=0}^{d-1}|i\rangle|i'\rangle$
for some orthonormal basis
$\{|i\rangle\}$ of $\mathbb{C}^{d}$ and some orthonormal
set $\{|i'\rangle\}$ of $\mathbb{C}^{d'}$.

Let $\mathcal{M}_{d\times d'}$ be the space of all $d$ by $d'$ complex matrices.
Then $\mathcal{M}_{d\times d'}$ is a Hilbert space equipped
with the inner product defined by $\langle A|B\rangle={\rm Tr}(A^{\dag}B)$
for any $A$, $B\in\mathcal{M}_{d\times d'}$.
$\{A_i: \ {\rm Tr}(A_i^{\dag}A_j)=\delta_{ij}, i=1,2,\dots,d_1d_2\}$
is called a Hilbert-Schmidt basis of $\mathcal{M}_{d\times d'}$.
There is a one-to-one relation between $\{|\psi_i\rangle\}$
and the Hilbert-Schmidt basis $\{A_i\}$ \cite{Guoyu2015jpa,Guoyu2015qip}:
\begin{widetext}
\begin{eqnarray}
|\psi_i\rangle=\sum_{k,l}a^{(i)}_{kl}|k\rangle|l'\rangle\in\mathbb{C}^{d}\otimes\mathbb{C}^{d'}
\Leftrightarrow A_i=[a_{kl}^{(i)}]\in\mathcal{M}_{d\times d'}, \ \
S_r(|\psi_i\rangle)={\rm rank}(A_i),\
\langle\psi_i|\psi_j\rangle={\rm Tr}(A_i^\dag A_j),
\label{relation}
\end{eqnarray}
\end{widetext}
where $\{|k\rangle\}$ and
$|l'\rangle$ are the standard computational bases of $\mathbb{C}^{d}$ and $\mathbb{C}^{d'}$,
respectively, and $S_r(|\psi_i\rangle)$ denotes the Schmidt number of $|\psi_i\rangle$.
For simplicity, we call a rectangular matrix a singular-value-1
matrix if its singular values are $\{1,1,\dots,1\}$.
Then $\sqrt{d}A_i$ is a $d\times d'$ singular-value-1 matrix
iff $|\psi_i\rangle$ is a maximally entangled pure state in $\mathbb{C}^{d}\otimes\mathbb{C}^{d'}$ and vice versa.
Especially, $\sqrt{d}A_i$ is a $d\times d$ unitary matrix iff
$|\psi_i\rangle$ is a maximally entangled pure state in $\mathbb{C}^{d}\otimes\mathbb{C}^{d}$.
For simplicity, we give the following definitions.

{\it Definition 1.~\cite{Bravyi,Chen}}  A set of states
$\{|\phi_i\rangle\in\mathbb{C}^{d}\otimes\mathbb{C}^{d'}:
i=1,2,\dots,n,n<dd'\}$ is called an
$n$-member UMEB if and only if

(i) $|\phi_i\rangle$, $i=1,2,\dots,n$, are maximally entangled;

(ii) $\langle\phi_i|\phi_j\rangle=\delta_{ij}$;

(iii) if $\langle\phi_i|\psi\rangle=0$ for all $i=1,2,\dots,n$,
then $|\psi\rangle$ cannot be maximally entangled.


{\it Definition 2.} Let $\Omega=\{A_i: i=1,2,\dots,d^2\}$
be a Hilbert-Schmidt basis in $\mathcal{M}_{d\times d}$.
$\Omega$ is called a unitary Hilbert-Schmidt basis (UB) of
$\mathcal{M}_{d\times d}$ if $A_i$s are unitary matrices.

It is clear that $\Omega=\{A_i: i=1,2,\dots,d^2\}$
is a UB iff $\{|\psi_i\rangle\}$
is a maximally entangled basis (MEB) of $\mathbb{C}^{d}\otimes\mathbb{C}^{d}$.

{\it Definition 3.} Let $\Omega=\{A_i: i=1,2,\dots,dd'\}$
be a Hilbert-Schmidt basis in $\mathcal{M}_{d\times d'}$.
$\Omega$ is called a singular-value-1 Hilbert-Schmidt basis (SV1B) of
$\mathcal{M}_{d\times d'}$ if $A_i$s are singular-value-1 matrices.

{\it Definition 4.} A set of $d\times d$ unitary matrices
$\{U_i: i=1,2,\dots,n,n<d^2\}$ is called an unextendible
unitary Hilbert-Schmidt basis (UUB) of $\mathcal{M}_{d\times d}$ if

i) ${\rm Tr}(U_i^{\dag}U_j)=\delta_{ij}$;

ii) if ${\rm Tr}(U_i^{\dag}X)=0$, $i=1$, 2, $\dots$, $n$, then $X$ is not unitary.

{\it Definition 5.} A set of $d\times d'$ ($d<d'$) singular-value-1 matrices
$\tilde{\Omega}=\{A_i: i=1,2,\dots,n,n<dd'\}$ is called an unextendible
singular-value-1 Hilbert-Schmidt basis (USV1B) of $\mathcal{M}_{d\times d}$ if

i) ${\rm Tr}(A_i^{\dag}A_j)=\delta_{ij}$;

ii) if ${\rm Tr}(A_i^{\dag}X)=0$, $i=1$, 2, $\dots$,
$n$, then $X$ is not a singular-value-1 matrix.

It is obvious that $\{U_i\}$ is a UUB of $\mathcal{M}_{d\times d}$
if and only if $\{|\psi_i\rangle\}$
is a UMEB of $\mathbb{C}^{d}\otimes\mathbb{C}^{d}$ while
$\{A_i\}$ is a USV1B of $\mathcal{M}_{d\times d'}$
if and only if $\{|\psi_i\rangle\}$
is a UMEB of $\mathbb{C}^{d}\otimes\mathbb{C}^{d'}$.

We will close this section with three lemmas which are necessary
in the proof of our main results in Secs. III and IV.
The Lemma 1 below is borrowed from \cite{Guoyu2015jpa}, which reveals that SV1B exists
in $\mathcal{M}_{d\times d'}$ for any $d$ and $d'$, $d<d'$.

{\it Lemma 1. \cite{Guoyu2015jpa}} MEB exists in $\mathbb{C}^{d}\otimes\mathbb{C}^{d'}$ for any $d$ and $d'$.

We remark here that, MEB is a complete basis of the space, that is,
it is a basis with any element is indeed a maximally entangled pure state.
The following lemmas can be easily checked.

{\it Lemma 2.} Let $\mathcal{M}_{d\times d}=\mathcal{L}\oplus\mathcal{L}^\bot$.
If $\{U_i: i=1,2,\dots,n,n<d^2\}$
is a UB of $\mathcal{L}$ and $\{V_i: i=1,2,\dots,m,m<d^2-n\}$
is a UUB of $\mathcal{L}^\bot$,
then $\{U_i: i=1,2,\dots,n,n<d^2\}\cup\{V_i: i=1,2,\dots,m,m<d^2-n\}$ is a UUB of $\mathcal{M}_{d\times d}$;
If $\{U_i: i=1,2,\dots,n,n<d^2\}$
is a UB of $\mathcal{L}$ and $\mathcal{L}^\bot$
contains no unitary matrix,
then $\{U_i: i=1,2,\dots,n,n<d^2\}$ is a UUB of $\mathcal{M}_{d\times d}$.

{\it Lemma 3.} Let $\mathcal{M}_{d\times d'}=\mathcal{L}\oplus\mathcal{L}^\bot$ with respect to
the Hilbert-Schmidt inner product and $\{A_i: i=1,2,\dots,n,n<d^2\}$
be a SV1B of $\mathcal{L}$. If $\mathcal{L}^\bot$ contain no singular-value-1 matrix,
then $\{A_i: i=1,2,\dots,n,n<d^2\}$ is a USV1B of $\mathcal{M}_{d\times d'}$.

\section{$d=d'$}

To construct a UMEB in any given bipartite state space, one always expects to
get one that consisting of fewer elements.
The following theorem is the main result of this section.
It improves the previous work in Ref. \cite{Wangyanling}
by giving fewer members of UMEB.

{\it Theorem 1.} If there is an $N$-member UMEB in $\mathbb{C}^{d}\otimes\mathbb{C}^{d}$,
then for any $q$, there is an $\tilde{N}$-member,$\tilde{N}=(qd)^2-q(d^2-N)$,
UMEB in $\mathbb{C}^{qd}\otimes\mathbb{C}^{qd}$.

{\it Proof.} Let
\begin{eqnarray*}
S_q=\left(\begin{array}{ccccc}
0&1&0&\cdots&0\\
0&0&1&\cdots&0\\
\vdots&\vdots&\vdots&\ddots&\vdots\\
0&0&0&\cdots&1\\
1&0&0&\cdots&0\end{array}\right),~~~~\\
\mathcal{L}_{q,k}^{(d)}=\{S_q^k\otimes X: X\in\mathcal{M}_{d\times d}\}, \ 1\leq k\leq q,
\end{eqnarray*}
then
\begin{eqnarray}
\mathcal{M}_{qd\times qd}=\bigcup\limits_{k=1}^q\mathcal{L}_{q,k}^{(d)}.
\end{eqnarray}
By Lemma 1, UB exists in $\mathcal{M}_{d\times d}$ for any $d$.
Let $\{U_i:i=1,\dots,d^2\}$ be a UB in $\mathcal{M}_{d\times d}$,
$\zeta_q=e^{\frac{2\pi}{q}{\rm i}}$, where ${\rm i}=\sqrt{-1}$,
and let
\begin{eqnarray*}
T_{q(j)}=\left(\begin{array}{ccccc}
1&\zeta_q^j&\zeta_q^{2j}&\cdots&\zeta_q^{j(q-1)}\\
1&\zeta_q^j&\zeta_q^{2j}&\cdots&\zeta_q^{j(q-1)}\\
\vdots&\vdots&\vdots&\ddots&\vdots\\
1&\zeta_q^j&\zeta_q^{2j}&\cdots&\zeta_q^{j(q-1)}\end{array}\right).
\end{eqnarray*}
We define
\begin{eqnarray}
U^{(k,j)}_{i}:=(S_q^k\circ T_{q(j)})\otimes U_i,
\end{eqnarray}
where $\circ$ denotes the Hadamard product, $j=0,1,\dots,q-1, k=1,\dots,q-1$.
It follows that
$\mathcal{U}_{q,d}:=\{U^{k,j}_{i}:i=1,\dots,d^2, j=0,1,\dots,q-1, k=1,\dots,q-1 \}$
forms a UB in
$\bigcup\limits_{k=1}^{q-1}\mathcal{L}_{q,k}^{(d)}$.
We now take a UUB of $\mathcal{M}_{d\times d}$, denote by $\{V_i:i=1,\dots,N<d^2\}$,
and let
\begin{eqnarray}
V_{i}^{(j)}:=(I_q\circ T_{q(j)})\otimes V_i,
\end{eqnarray}
where $I_q=S_q^q$ is the $q\times q$ identity matrix, $i=1,\dots,N,j=0,1,\dots,q-1$.
We assert that
\begin{eqnarray*}
\mathcal{V}_{q,d}:=\{V_{i}^{(j)}:i=1,\dots,N,j=0,1,\dots,q-1\}
\end{eqnarray*}
is a UUB of $\mathcal{L}_{q,q}^{(d)}$.
Note that any matrix orthogonal to $\bigcup\limits_{k=1}^{q-1}\mathcal{L}_{q,k}^{(d)}$
is contained in $\mathcal{L}_{q,q}^{(d)}$. 
If $\mathcal{V}_{q,d}$ is not a UUB, to reach a contradiction, we assume that
$W=W_1\oplus W_2\oplus \cdots\oplus W_q$ is a unitary matrix in $\mathcal{L}_{q,q}^{(d)}$ satisfying
${\rm Tr}(W^\dag V_{j}^{(i)})=0$ for any $i=1$,$\dots$, $N$, and $j=0$, 1, $\dots$, $q-1$.
We write ${\rm Tr}(W_j^\dag V_i)=\alpha_{ij}$, it turns out that
\begin{eqnarray}
\left(\begin{array}{ccccc}
1&1&1&\cdots&1\\
1&\zeta_q&\zeta_q^{2}&\cdots&\zeta_q^{q-1}\\
1&\zeta_q^2&\zeta_q^{4}&\cdots&\zeta_q^{2(q-1)}\\
\vdots&\vdots&\vdots&\ddots&\vdots\\
1&\zeta_q^{q-1}&\zeta_q^{2(q-1)}&\cdots&\zeta_q^{(q-1)^2}\end{array}\right)
\left(\begin{array}{c}
\alpha_{i1}\\
\alpha_{i2}\\
\alpha_{i3}\\
\vdots\\
\alpha_{iq}
\end{array}\right)=0\label{relation2}
\end{eqnarray}
holds for any $i$, $1\leq i\leq N$.
Eq.~(\ref{relation2}) leads to $\alpha_{ij}=0$ for any $i$ and $j$, which is impossible since
$\{V_i:i=1,\dots,N<d^2\}$ is unextendible.
We now conclude from Lemma 2 that
\begin{eqnarray}
\mathcal{U}_{q,d}\bigcup \mathcal{V}_{q,d}
\end{eqnarray}
is a UUB of $\mathcal{M}_{qd\times qd}$.
Then we can get a $(qd)^2-q(d^2-N)$-member UMEB corresponding to $\mathcal{U}_{q,d}\bigcup \mathcal{V}_{q,d}$ by Eq.~(\ref{relation}).
\hfill$\blacksquare$

Comparing with the $(qd)^2-(d^2-N)$-member UMEB in $\mathbb{C}^{qd}\otimes \mathbb{C}^{qd}$
in \cite{Wangyanling},
we get a UMEB consisting of less number of elements.
In addition, Theorem 1 reveals that we can obtain UMEB from not only the UMEB in $\mathbb{C}^{d}\otimes \mathbb{C}^{d}$
but also the UMEB in $\mathbb{C}^{q}\otimes \mathbb{C}^{q}$, and moreover, the two kinds of UMEBs have the same number of members.
But they are not equivalent to each other in general
(here, two sets of UMEB $\{|\psi_i\rangle\}_{i=1}^n$ and $\{|\phi_i\rangle\}_{i=1}^n$
are called equivalent if there exists a permutation $\pi\in\mathcal{S}_n$, unitary matrices $U$ and $V$ such that
$U\otimes V|\psi_i\rangle=|\phi_{\pi(i)}\rangle$ for $i=1$, $\dots$, $n$,
where $\mathcal{S}_n$ denotes the permutation group of $n$ elements \cite{Wangyanling}).
We illustrate it with the case of $\mathbb{C}^{12}\otimes \mathbb{C}^{12}$.
Observe that, $\{|\psi_i\rangle \}_{i=1}^n$ and $\{|\phi_i\rangle\}_{i=1}^n$
are equivalent iff the corresponding UUBs $\mathcal{U}_{\psi}$ and $\mathcal{V}_{\phi}$
admit $UU_iU_jU^\dag=V_{\pi(i)}V_{\pi(j)}$ for any $U_i\in\mathcal{U}_{\psi}$
and $V_i\in\mathcal{V}_{\phi}$ \cite{Wangyanling}. This reveals that $U_iU_j$ and $V_{\pi(i)}V_{\pi(j)}$
have the same eigenvalues if the two UMEBs are equivalent.
On one hand, any element in $\mathcal{V}_{4,3}$ has a eigenvalue $e^{{\rm i}\theta}$ with infinite order
(here, the order of a complex number, $\lambda$, $|\lambda|=1$, is the least nature number $n$ such that $\lambda^n=1$),
where $\theta$ satisfying $\cos\theta=-\frac{7}{8}$ \cite{Wangyanling}.
On the other hand, any element in $\mathcal{V}_{3,4}$ does not contain eigenvalue with infinite order.
Thus the two UMEBs in $\mathbb{C}^{12}\otimes \mathbb{C}^{12}$
(i.e., the two UMEBs corresponding to $\mathcal{V}_{4,3}$ and
$\mathcal{V}_{3,4}$ respectively) based on our scenario can not be equivalent.
We also remark here that, let $\mathcal{L}_{q,d}$ be the space with $\mathcal{U}_{q,d}\cup \mathcal{V}_{q,d}$ as its UB,
then $\mathcal{L}_{q,d}^\bot$ does not contain unitary matrix.
In what follows, we give an example to illustrate Theorem 1.

{\it Example 1.} For any $d=2p$, if there is an $m$-member UUB
in $\mathcal{M}_{p\times p}$, then there is a $(2p^2+2m)$-member UMEB in $\mathbb{C}^d\otimes\mathbb{C}^d$.
Let $\{U_1,U_2,\dots,U_{p^2}\}$ be a UB of $\mathcal{M}_{p\times p}$
and let $\{V_1,V_2,\dots,V_m\}$ be a UUB of $\mathcal{M}_{p\times p}$.
Taking
\begin{eqnarray*}
\check{V}_i=
\left(\begin{array}{cc}
0_p&V_i\\
V_i&0_p\end{array}\right),\
\check{V}^-_i=\left(\begin{array}{cc}
0_p&-V_i\\
V_i&0_p\end{array}\right),
\end{eqnarray*}
then, by Theorem 1,
\begin{eqnarray*}
\{\pm U_i\oplus U_i\}_{i=1}^{p^2} \cup \{ \check{V}_i\}_{i=1}^m \cup \{ \check{V}^-_i\}_{i=1}^m
\end{eqnarray*}
is a $(2p^2+2m)$-member UUB in $\mathcal{M}_{d\times d}$.

We may construct UMEB from other ways.
For the space of $\mathcal{M}_{d\times d}$ with
$d=s+t$, $2\leq s\leq t$, we let
\begin{eqnarray*}
\mathcal{L}_{s\oplus t}^{(d)}:=
\left\{\left(\begin{array}{cc}
A&0 \\
0&B\end{array}\right):A\in\mathcal{M}_{s\times s}, B\in\mathcal{M}_{t\times t}\right\},\\
\mathcal{L}_{s,t}^{(d)}:=
\left\{\left(\begin{array}{cc}
0&X \\
Y&0\end{array}\right):X\in\mathcal{M}_{s\times t}, Y\in\mathcal{M}_{t\times s}\right\},
\end{eqnarray*}
then
\begin{eqnarray}
\mathcal{M}_{d\times d}=\mathcal{L}_{s\oplus t}^{(d)}\oplus\mathcal{L}_{s,t}^{(d)}.
\end{eqnarray}
By Lemma 2, it is clear that if $s<t$, then any UB of $\mathcal{L}_{s\oplus t}^{(d)}$ (if it exists)
is a UUB of $\mathcal{M}_{d\times d}$.

{\it Proposition 1.} For any $d=s+t$, $2\leq s< t$, if $\Gamma$ is
a $(s^2+t^2)$-member MEB corresponding to the subspace $\mathcal{L}_{s\oplus t}^{(d)}$,
then $\Gamma$ is a $(s^2+t^2)$-member UMEB in $\mathbb{C}^d\otimes\mathbb{C}^d$.

However, it is hard to know whether there exists UB in $\mathcal{L}_{s\oplus t}^{(d)}$ for $s<t$.
The case of $s=t$ can be easily checked.
In fact,
let $\{U_1,U_2,\dots,U_{s^2}\}$ be a UB of $\mathcal{M}_{s\times s}$
then $\{\pm U_i\oplus U_i: i=1,2,\dots,s^2\}$ is a UB of $\mathcal{L}_{s\oplus s}^{(d)}$.
If it is true for $s<t$ we may get new UMEBs with much less members
and the existence problem of UMEB would be settled down completely.

\section{$d<d'$}

We now discuss the case of $d<d'$.
We consider the case of $2\otimes 3$ at first.
we denote
\begin{eqnarray*}
\left(\begin{array}{cccc}
*&*&* \\
*&*&*\end{array}\right)=
\left(\begin{array}{cccc}
*&*&0 \\
*&*&0  \end{array}\right)
\oplus
\left(\begin{array}{cccc}
0&0&* \\
0&0&*  \end{array}\right)
\end{eqnarray*}
by
\begin{eqnarray}
\mathcal{M}_{2\times3}=\mathcal{L}_{2\times2}^{(2,3)}\oplus\mathcal{L}_{3-1}^{(2,3)}.
\end{eqnarray}
By Lemma 1, SV1B exists in any matrix space. Thus, any SV1B of
the subspace $\mathcal{L}_{2\times2}^{(2,3)}$
is a USV1B of $\mathbb{C}^2\otimes\mathbb{C}^3$ since there is
no singular-value-1 matrix in $\mathcal{L}_{3-1}^{(2,3)}$.
Therefore, by Eq.~(\ref{relation}) and Lemma 3, the vectors
corresponding to SV1B of the subspace $\mathcal{L}_{2\times2}^{(2,3)}$ is a UMEB.
For the $2\otimes 4$ space,
\begin{eqnarray*}
\left(\begin{array}{cccc}
*&*&*&* \\
*&*&*&*\end{array}\right)=
\left(\begin{array}{cccc}
*&*&*&0 \\
*&*&*&0  \end{array}\right)
\oplus
\left(\begin{array}{cccc}
0&0&0&* \\
0&0&0&*  \end{array}\right).
\end{eqnarray*}
It is clear that any MEB corresponding to the subspace
$\left(\begin{array}{cccc}
*&*&*&0 \\
*&*&*&0  \end{array}\right)$ is a UMEB in $\mathbb{C}^2\otimes\mathbb{C}^3$.
In fact, for $\mathbb{C}^2\otimes\mathbb{C}^{d'}$ with any $d'$,
we denote
\begin{eqnarray*}
\left(\begin{array}{ccc}
*&\cdots&* \\
*&\cdots&*\end{array}\right)=
\left(\begin{array}{cccc}
*&\cdots&*&0 \\
*&\cdots&*&0  \end{array}\right)
\oplus
\left(\begin{array}{cccc}
0&\cdots&0&* \\
0&\cdots&0&*  \end{array}\right)
\end{eqnarray*}
by
\begin{eqnarray}
\mathcal{M}_{2\times d'}=\mathcal{L}_{2\times (d'-1)}^{(2,d')}\oplus\mathcal{L}_{d'-1}^{(2,d')}.
\end{eqnarray}
Then any MEB of the first subspace $\mathcal{L}_{2\times (d'-1)}^{(2,d')}$
is a UMEB in $\mathbb{C}^2\otimes\mathbb{C}^{d'}$.

Hereafter,
we let
\begin{eqnarray}
\mathcal{L}_{d\times(d'-i)}^{(d,d')}=\left(\begin{array}{cccccc}
*&\cdots&*&0&\cdots&0 \\
\vdots&\ddots&\vdots&\vdots&\ddots&\vdots \\
*&\cdots&*&0&\cdots&0
\end{array}\right)
\end{eqnarray}
(all the entries of the last $i$ columns are zeros, $1\leq i<d$) and let $\Gamma_{d\times(d'-i)}^{(d,d')}$
be the MEB of the subspace $\mathcal{L}_{d\times(d'-i)}^{(d,d')}$, we have

{\it Theorem 2.} i) If $d'\geq 2d$, then $\Gamma_{d\times(d'-i)}^{(d,d')}$ is a
$d(d'-i)$-member UMEB in $\mathbb{C}^d\otimes \mathbb{C}^{d'}$
for any $1\leq i<d$; ii) If $d'=d+r$, $1\leq r<d$, then $\Gamma_{d\times(d'-i)}^{(d,d')}$ is a
$d(d'-i)$-member UMEB in $\mathbb{C}^d\otimes \mathbb{C}^{d'}$
for any $1\leq i\leq r$.

We also have other method to get UMEB.
If $d'\geq2d$, we consider the decomposition
\begin{eqnarray}
\mathcal{M}_{d\times d'}=\mathcal{L}_{d\times(d'-d)}^{(d,d')}\oplus\mathcal{L}_{d'-d}^{(d,d')}.
\end{eqnarray}
We denote by $\Gamma_{d\times(d'-d)}^{(d,d')}$ the MEB
corresponding to $\mathcal{L}_{d\times(d'-d)}^{(d,d')}$
and by $\tilde{\Gamma}_{d'-d}^{(d,d')}$ the UMEB
corresponding to $\mathcal{L}_{d'-d}^{(d,d')}$ (if it exists).
It turns out that

{\it Proposition 2.} If $d'\geq2d$ and there exists
a UMEB $\tilde{\Gamma}_{d'-d}^{(d,d')}$ corresponding to $\mathcal{L}_{d'-d}^{(d,d')}$,
then
$\Gamma_{d\times(d'-d)}^{(d,d')}\bigcup\tilde{\Gamma}_{d'-d}^{(d,d')}$
is a UMEB of $\mathbb{C}^d\otimes \mathbb{C}^{d'}$.

Observing that $\mathcal{L}_{d'-d}^{(d,d')}\cong\mathcal{M}_{d\times d}$,
so $\tilde{\Gamma}_{d'-d}^{(d,d')}$ can be viewed as a
UMEB in $\mathcal{M}_{d\times d}$ (if it exists).
For clarity, we list some examples below.

{\it Example 2.} By Theorem 2, it is obvious that
\begin{eqnarray*}
|\phi_1\rangle&=&\frac{1}{\sqrt{2}}(|0\rangle|0'\rangle+|1\rangle|1'\rangle),\\
|\phi_2\rangle&=&\frac{1}{\sqrt{2}}(|0\rangle|0'\rangle-|1\rangle|1'\rangle),\\
|\phi_3\rangle&=&\frac{1}{\sqrt{2}}(|0\rangle|1'\rangle+|1\rangle|0'\rangle),\\
|\phi_4\rangle&=&\frac{1}{\sqrt{2}}(|0\rangle|1'\rangle-|1\rangle|0'\rangle),
\end{eqnarray*}
constitute a UMEB in $\mathbb{C}^2\otimes \mathbb{C}^3$.
In general, in a $d\otimes d'$ ($d< d'$) system,
let
\begin{eqnarray*}
|\Omega_{0,0}\rangle=\frac{1}{\sqrt{d}}\sum\limits_{i=0}^{d-1}|i\rangle|i'\rangle
\end{eqnarray*}
and
\begin{eqnarray*}
\tilde{W}_{m,n}|i'\rangle=\xi_d^{mi}|(i-n)'\rangle,
\label{weyloperation3}
\end{eqnarray*}
where
 $|(i-n)'\rangle\equiv|(i-n)'\oplus(d-k)'\rangle$ (here $(i-n)'\oplus(d-k)'$ means $(i-n-k+d)'$ mod $(d-k)'$), $k'<d$,
whenever $d'\geq 2d$, and $|(i-n)'\rangle\equiv|(i-n)'\oplus(d-l)'\rangle$
(here $(i-n)'\oplus(d-l)'$ means $(i-n-l+d)'$ mod $(d-l)'$), $1\leq l'\leq r$,
whenever $d'=d+r$, $1\leq r<d$.
Then
\begin{eqnarray}
|\tilde{\Omega}_{m,n}\rangle=( \mathbbm{1} \otimes \tilde{W}_{m,n})|\Omega_{0,0}\rangle
\label{meb2}
\end{eqnarray}
with $0\leq m \leq d-1$ and
\begin{eqnarray*}
\left\{\begin{array}{ll}
0\leq n\leq d'-k' & {\rm whenever}\ d'\geq 2d, \\
0\leq n\leq d'-l' &{\rm whenever}\ d'=d+r \ {\rm and}\ 1\leq r<d,
\end{array}\right.
\end{eqnarray*}
induce a UMEB in $\mathbb{C}^d\otimes \mathbb{C}^{d'}$.

{\it Example 3.} In $\mathbb{C}^3\otimes \mathbb{C}^6$, we can obtain a 15-member UMEB from the 6-member UMEB in
$\mathbb{C}^3\otimes \mathbb{C}^3$ proposed in \cite{Bravyi}.
Taking
$|\psi_{1,2}\rangle=\frac{1}{a}(|0'\rangle\pm b|1'\rangle)$,
$|\psi_{3,4}\rangle=\frac{1}{a}(|1'\rangle\pm b|2'\rangle)$,
$|\phi_{5,6}\rangle=\frac{1}{a}(|2'\rangle\pm b|0'\rangle)$,
where $b=(1+\sqrt{5})/2$, $a=\sqrt{1+b^2}$, and defining
\begin{eqnarray*}
U_j=I-(1-e^{{\rm i}\theta})|\psi_j\rangle\langle\psi_j|,\quad j=1, 2, \dots, 6,
\end{eqnarray*}
where $\cos\theta=-\frac{7}{8}$, we let
\begin{eqnarray*}
|u_j\rangle:=(I\otimes U_j)|\Phi^+\rangle,\quad j=1,2,\dots,6,
\end{eqnarray*}
where $|\Phi^+\rangle=\frac{1}{\sqrt{3}}\sum_{i=0}^2|i\rangle|i'\rangle$
(namely, $\{|u_j\rangle: j=1,2,\dots,6,\}$ constitute a 6-member UMEB in $\mathbb{C}^3\otimes \mathbb{C}^3$~\cite{Bravyi}).
Let
\begin{eqnarray*}
|\phi_1\rangle&=&\frac{1}{\sqrt{3}}(|0\rangle|0'\rangle+|1\rangle|1'\rangle+|2\rangle|2'\rangle),\\
|\phi_2\rangle&=&\frac{1}{\sqrt{3}}(|0\rangle|0'\rangle+\xi|1\rangle|1'\rangle+\xi^2|2\rangle|2'\rangle),\\
|\phi_3\rangle&=&\frac{1}{\sqrt{3}}(|0\rangle|0'\rangle+\xi^2|1\rangle|1'\rangle+\xi^4|2\rangle|2'\rangle),\\
|\phi_4\rangle&=&\frac{1}{\sqrt{3}}(|0\rangle|4'\rangle+|1\rangle|5'\rangle+|2\rangle|3'\rangle),\\
|\phi_5\rangle&=&\frac{1}{\sqrt{3}}(|0\rangle|4'\rangle+\xi|1\rangle|5'\rangle+\xi^2|2\rangle|3'\rangle),\\
|\phi_6\rangle&=&\frac{1}{\sqrt{3}}(|0\rangle|4'\rangle+\xi^2|1\rangle|5'\rangle+\xi^4|2\rangle|3'\rangle),\\
|\phi_7\rangle&=&\frac{1}{\sqrt{3}}(|0\rangle|5'\rangle+|1\rangle|3'\rangle+|2\rangle|4'\rangle),\\
|\phi_8\rangle&=&\frac{1}{\sqrt{3}}(|0\rangle|5'\rangle+\xi|1\rangle|3'\rangle+\xi^2|2\rangle|4'\rangle),\\
|\phi_9\rangle&=&\frac{1}{\sqrt{3}}(|0\rangle|5'\rangle+\xi^2|1\rangle|3'\rangle+\xi^4|2\rangle|4'\rangle),
\end{eqnarray*}
then, $\{|u_j\rangle: j=1,2,\dots,6,\}\cup\{|\phi_i\rangle:i=1,\dots,9\}$
is a 15-member UMEB in $\mathbb{C}^3\otimes \mathbb{C}^6$ by Proposition 2.

Comparing with all the previous work in \cite{Chen,Limaosheng},
one can easily find out that all the results there are special cases of Theorem 2 and Proposition 2.
By now, in $\mathbb{C}^d\otimes \mathbb{C}^{d'}$,
there are at least $d$ kinds of ways to construct UMEB if $d'\geq 2d$ for any $d>2$
and there are at least $r$ sets of UMEB with different number of elements if $d'=d+r$, $r\geq1$, $d>2$.

\section{Conclusions}

This brief report has proposed an easy way to construct UMEB
in $\mathbb{C}^d\otimes\mathbb{C}^{d'}$ for any $d$ and $d'$ via the approach of
the associated matrix space decomposition. A more clear structure of the UMEB
is presented.
The MEB of some subspace happens to be the UMEB of the whole space.
In addition, in $\mathbb{C}^{qd}\otimes\mathbb{C}^{qd}$,
the members of the UMEB from our method is less than the known one before.
Together with the scenario in \cite{Wangyanling}, we now have four different
UMEB in $\mathbb{C}^{qd}\otimes\mathbb{C}^{qd}$ ($qd>6$).
For the $d\otimes d'$ case with $d<d'$, the protocol here
covers all the previous work as special cases.

In fact, all of the methods used in constructing UMEB can be regarded as some special
space decompositions of the associated matrix space by relation in Eq.~(\ref{relation}):
If the whole space can be decomposed as an orthogonal direct sum of subspaces
with one has UB while the other one contains no unitary matrix, then the UB is a UUB of
the whole space.
Therefore the key point comes down to find such a subspace.
The structure of UMEB is more clear under the subspace decomposition.
So now the remain open existing problem is reduced to find a UB of the subspace
associated with this kind of decomposition of
$\mathbb{C}^{p}\otimes\mathbb{C}^{p}$
whenever $p$ is a prime.

\begin{acknowledgements}

This
work was completed while the author was visiting the Institute of Quantum Science and
Technology of the University of Calgary during the academic year 2016-2017 under the
support of China Scholarship Council.
The author thanks Professor
Christoph Simon and Professor Alexander Lvovsky for their hospitality.
The author is supported by the National Natural Science Foundation of China under Grant No. 11301312.
\end{acknowledgements}



\end{document}